\documentclass{caosp306}
\usepackage{graphicx}
\usepackage{natbib}
\articleNo{123}
\pubyear{}
\volume{}
\volnumber{}
\firstpage{1}
\received{July 31, 2019}
\accepted{ }
\def\BibTeX{{\rm B\kern-.05em{\sc i\kern-.025em b}\kern-.08em
             T\kern-.1667em\lower.7ex\hbox{E}\kern-.125emX}}

\begin{document}

\hauthor{D.V.\,Kozlova, A.V.\,Moiseev and A.A.\,Smirnova}

\title{Extended ionized-gas structures in Seyfert 2 galaxy Mrk~78}

\author{
        D.V.\,Kozlova \inst{1,} \inst{2} 
      \and 
        A.V.\,Moiseev \inst{2}   
      \and 
        A.A.\,Smirnova \inst{2}
       }

\institute{
           Ural Federal University, 620002, Russia
         \and 
           Special Astrophysical Observatory of the Russian Academy of Sciences, 369167, Russia 
          }
                 
\date{March 8, 2003}

\maketitle

\begin{abstract}
Search for and study of extended emission-line regions (EELRs) related to AGN in early-type galaxies is interesting to probe the history of nuclear ionization activity and also to understand the process of external gas accretion. In this work, we present observations of the EELR in Mrk~78 obtained at the 6-m  Russian telescope using the long-slit and 3D spectroscopy methods. We show that ionized-gas clouds at the 12-16\,kpc projected distances from the nucleus are ionized by the AGN radiation. Also we have checked if the galaxy appearing in the optical images in the immediate neighbourhood of Mrk~78 near the external clouds is a dwarf companion  or a part of a tidal structure. However, the spectrum of this galaxy, SDSS J074240.37+651021.4, obtained at the 6-m telescope corresponds to the distant background galaxy.

\end{abstract}

\section{Introduction}
\label{intr}

According to the unified model of active galactic nuclei (AGN), the radiation of the central accretion engine comes out in  broad axisymmetrical cones collimated by the circumnuclear obscuring torus. The impact of AGN ionizing radiation together with radio jet  mechanical energy  on the surrounding interstellar medium appears in the shape of ionizing cones from hundreds pc to a few kpc in typical size.  In some fraction of Seyfert galaxies,  extended emission-line regions (EELRs) are also detected at radial distances of tens of kpc. EELRs can be used to 
measure the AGN energetic output at both time and spatial scales and also to study the gas distribution in the galactic environment. Indeed, a significant faction of EELR traces the off-plane gas related with tidal debris \citep{Keel2012} or even the gas in a companion galaxy \citep{Keel2019}. One of the  intriguing cases is the well-known AGN  Mrk~6: deep imaging and spectroscopic observations revealed an extended system of gaseous filaments up to 40 kpc from the nucleus accreted by the Seyfert galaxy \citep{Smirnova2018}. Therefore, the detailed study of EELRs is topical even around well-studied nearby active galaxies. 

\begin{figure}[h]
\centerline{
\includegraphics[width=0.55\textwidth]{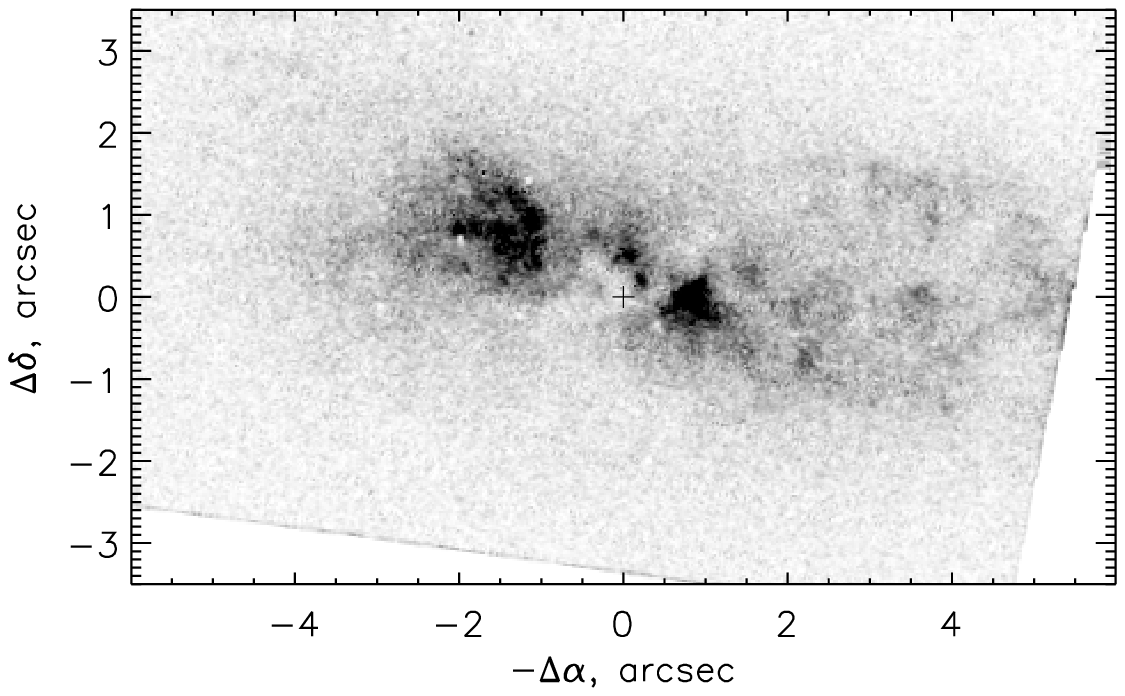}
\includegraphics[width=0.35\textwidth]{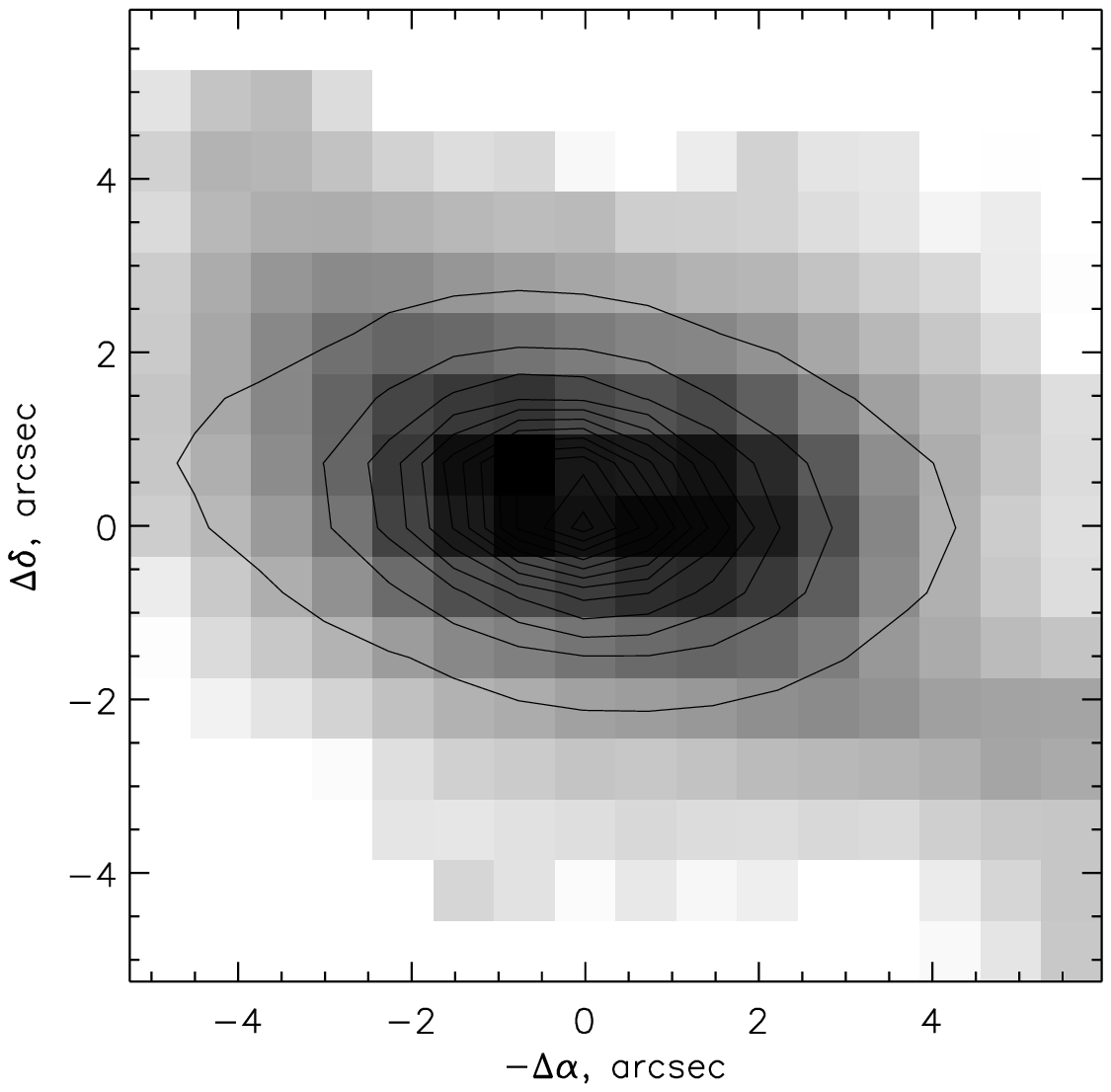}
}
\caption{Left: the HST/FOC image with the F520M filter, shows the  [OIII]  emission \citep[see][]{Fischer2011}; right: the MPFS image in the [OIII]${\lambda}5007$ line with the superimposed continuum isophotes.}
\label{MPFS}
\end{figure}

The Seyfert 2 galaxy Mrk~78 is a good illustration of the picture described above. The spectrophotometric properties of the inner ($r<4-5''$, or 3--4\,kpc) elongated emission-line structure studied at the HST \citep{2004AJ....127..606W,2006MNRAS.372..961K} can be explained in terms of jet-cloud interaction together with the mass outflow and a cone of UV-radiation emerging from the Sy nucleus \citep{Fischer2011}, see the corresponding [OIII] HST image in Fig.~\ref{MPFS}. Ground-based  narrow-band images obtained by \citet{DeRobertis1987}  and \citet{Pedlar1989} also reveal a weak extended ionized-gas emission outside the central region with an  extensive asymmetric structure at $r>15''$ in the south-west. The presence of the gas emitting  in  [OIII] in the western direction up to 12-14\,$''$ was  confirmed in spectroscopic observations by \citet{Unger1987}. Also,   \citet{AfanasievSilchenko1991} detected the [OIII] line emission even at $r=18-20''$. \citet{Pedlar1989} have shown that the weak emitting narrow-line region extending out to 10 kpc to the west is consistent with AGN ionizing cones, however, the actual ionization  state of the most distant part of the EELR is still unknown. Mrk~78 was included in the sample of the confirmed EELR selected from the Galaxy Zoo survey, but  the follow-up spectra were obtained along  $PA=90\,\degr$\citep{Keel2012}, therefore, the properties of the southwestern external emission knots were not measured.    

In this paper, we analysed the  large-scale ionized gas distribution and spectral properties of the most distant parts of the EELR in Mrk~78 derived from new long-slit and 3D spectroscopic observations at the  6-m telescope of Special Astrophysical Observatory of Russian Academy of Sciences (SAO RAS). The gas and stellar kinematics will be considered in the forthcoming paper. In the present paper, we accepted the distance to the galaxy as $165$\,Mpc with a corresponding scale of 0.80 \,kpc/$''$ (according to  NED\footnote{http://ned.ipac.caltech.edu/}).

\section{Observations and data reduction}
\label{obs}

Observations were carried out at the prime focus of the 6-m SAO RAS telescope with the multi-mode focal reducers: SCORPIO \citep{Afanasiev2005} and SCORPIO-2 \citep{Afanasiev2011}. We used the 3D spectroscopy with the Fabry-Perot interferometer (FPI), long-slit spectroscopy, and direct-image modes. Also, we used the data obtained with the integral-field Multi-Pupil Field Spectrograph \citep{2001sdcm.conf..103A}. The log of observations and parameters of the used instruments are listed in the Table\,\ref{t1}. 

SCORPIO and SCORPIO-2 both have the $6\farcm1$ fields of view with the 0.36-arcsec pixel size. The slit has the $6\farcm1$ length and the $1''$ width. The scanning FPI mapped the spectral region around the redshifted [OIII]${\lambda}4959$ emission line. We made the binning of these data to reach a higher signal-to-noise ratio  in the resulting data cube with 0.7\,$''$/px. The MPFS data cube had the 16x16 arcsec$^2$ field-of-view centered at the nucleus with the $1''/$spaxel spatial scale and covered a wide spectral range. 

Calibration of the SCORPIO-2 and MPFS spectra into the absolute energy flux units was carried out using spectrophotometric standard stars observed in the same night as Mrk~78. The data reduction was performed in a  standard way (see \citet{Smirnova2018} for references on the software and algorithms).  

Figure~\ref{fields} shows the maps derived from the Voight fitting of the [OIII] spectra in the FPI data cube: the monochromatic image, the velocity field, and the velocity dispersion map free from instrumental effects. The emission-line fluxes in the MPFS and long-slit data were derived from the single-component Gaussian fitting. The map of the [OIII] emission in the circumnuclear region according to the MPFS data is shown in Fig.~\ref{MPFS}. For the internal region, where well-known multi-component  lines were observed \citep[see][and references therein]{AfanasievSilchenko1991}, only a brighter peak was fitted in the MPFS and FPI data.

\begin{table}[t]
\small
\begin{center}
\caption{Log of observations at the SAO RAS telescope}
\label{t1}
\begin{tabular}{llllll}
\hline\hline
Date  & Instrument/mode & Exp. time, s & Sp. range & FWHM & Seeing\\
\hline
14/15 Mar 2007 & SCORPIO/FPI  & 32$\times$250 & [OIII]${\lambda}4959$ & 1.2\,\AA& 1\,\farcs2 \\
22/23 Oct 2008 & SCORPIO/DI  & 5$\times$120 & R$_C$-filter & & 1\,\farcs2 \\
03/04 Feb 2008 & MPFS  & $6\times1200$ & 3650--5900\,\AA & 5\AA& 1\,\farcs3 \\
03/04 Feb 2008 & MPFS  & $6\times1200$ & 4300--7380\,\AA & 5\AA& 1\,\farcs3 \\
12/13 Feb 2019 & SCORPIO-2/LS & 3x900 & 3650--8530\,\AA  & 7\AA & 2\,\farcs7  \\
\hline\hline
\end{tabular}
\end{center}
\end{table}

\section{External gas clouds}
\label{prop}

\begin{figure}[h]
\centerline{
\includegraphics[height=0.43\textwidth]{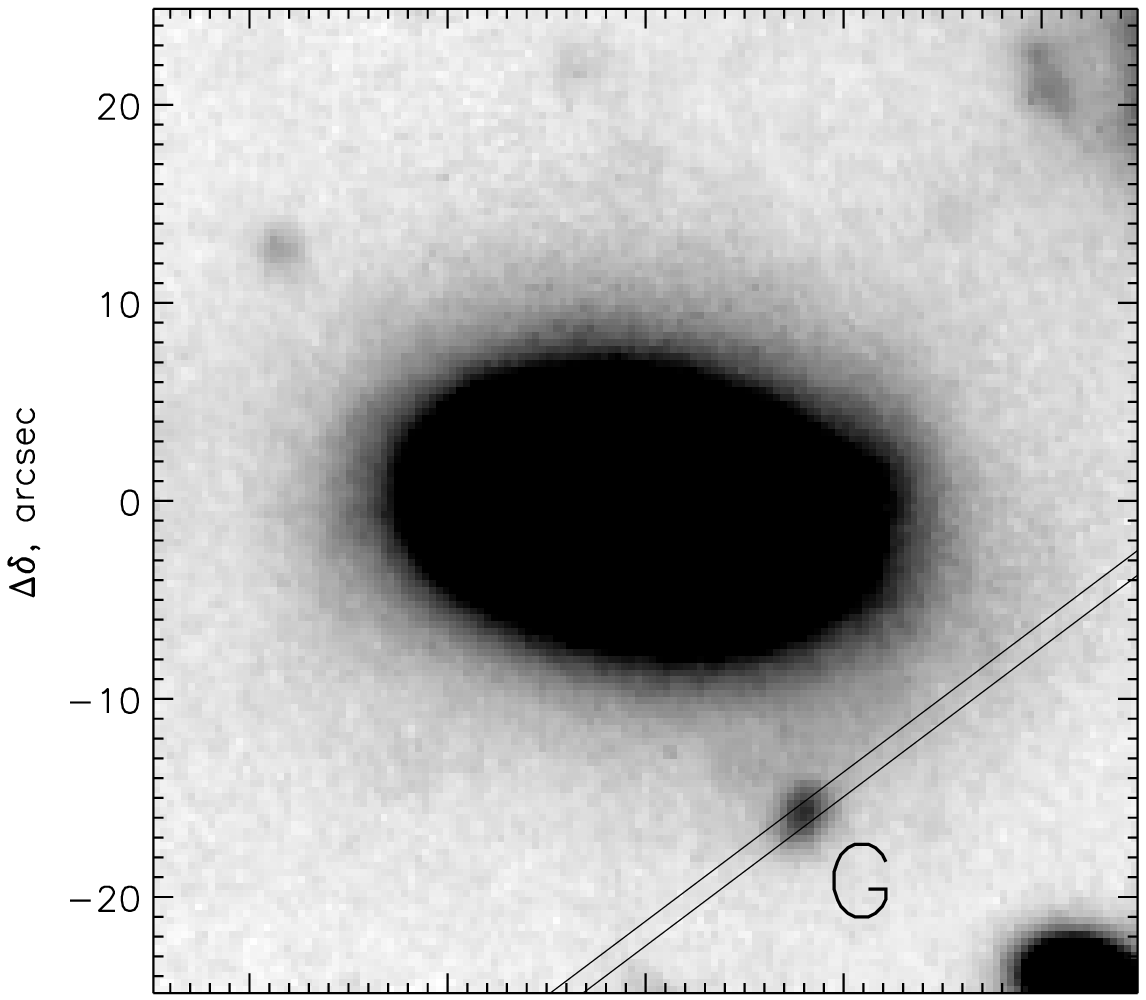}
\includegraphics[height=0.43\textwidth]{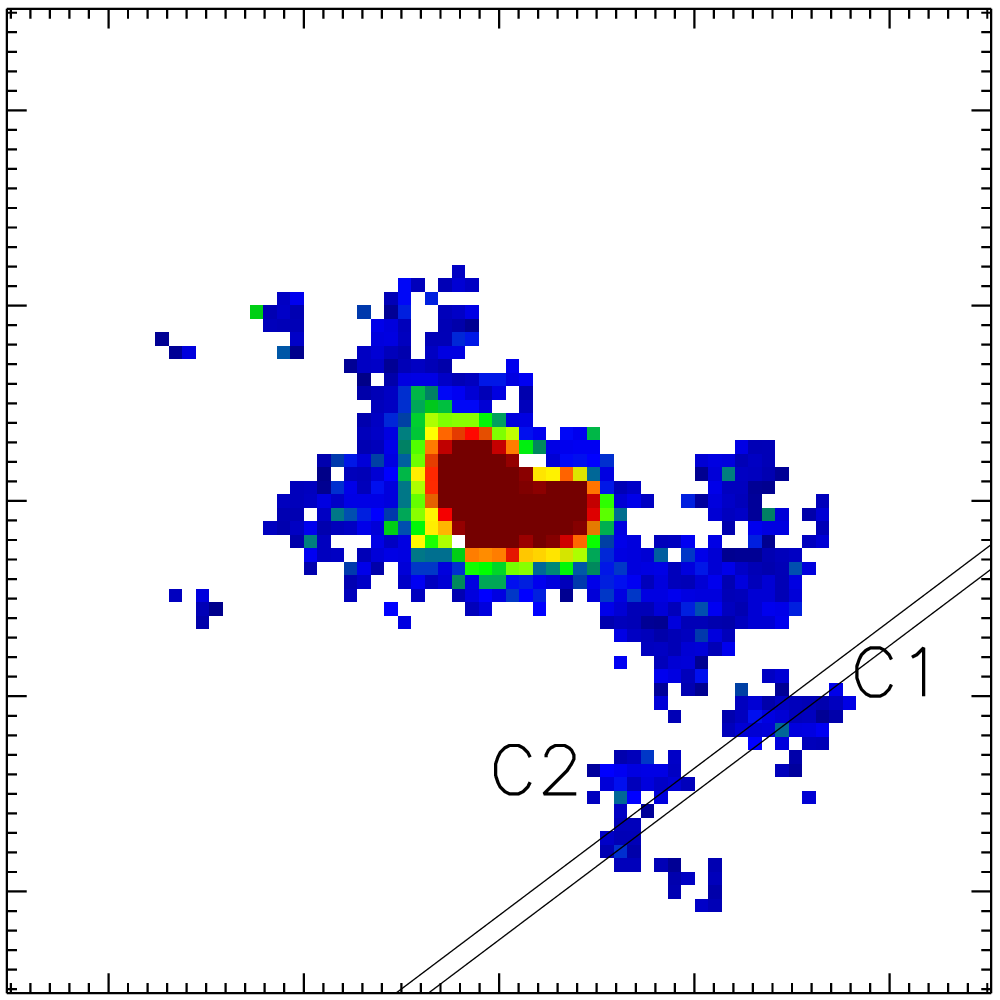}
}
\centerline{\includegraphics[height=0.53\textwidth]{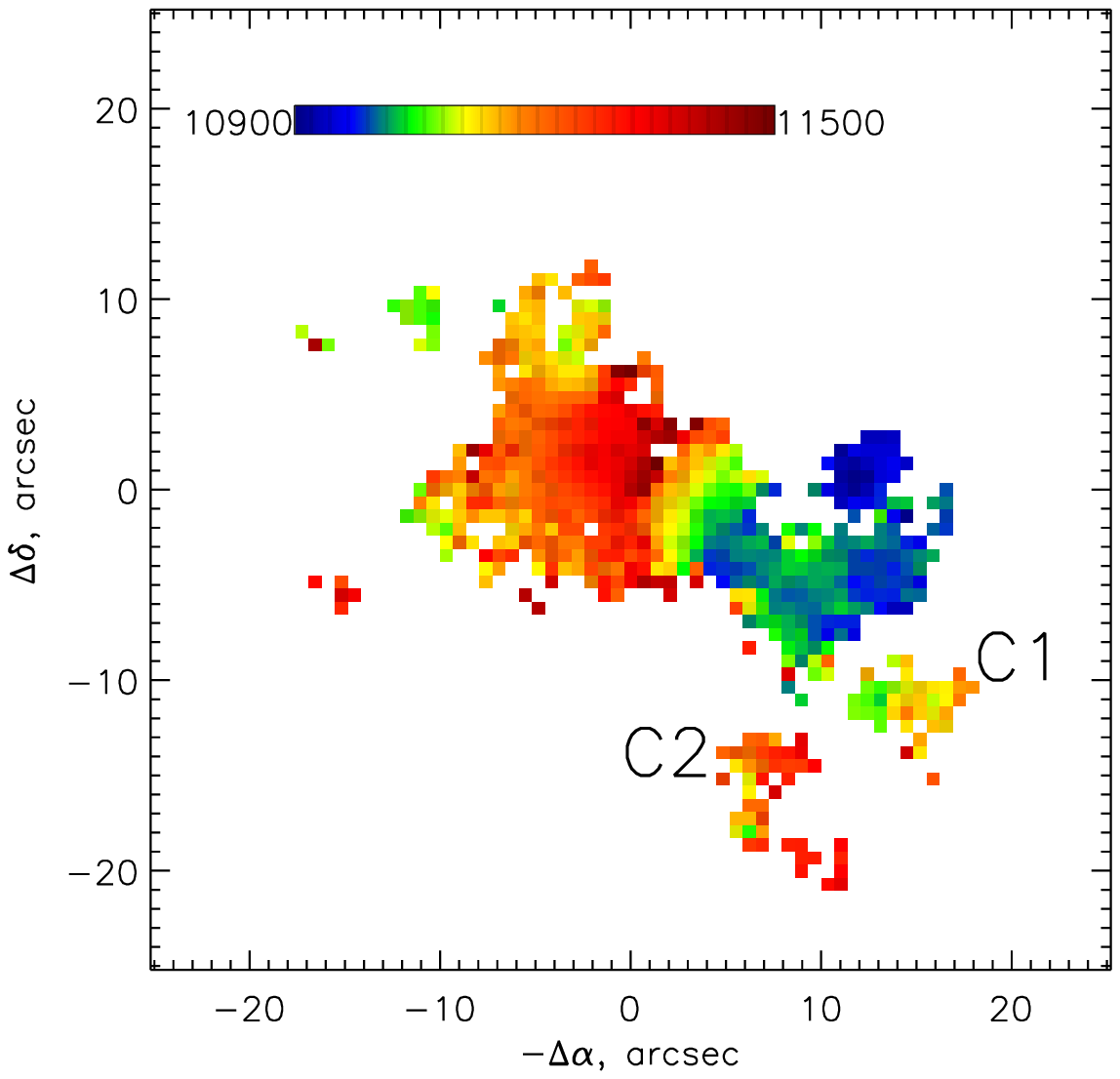}
\includegraphics[height=0.53\textwidth]{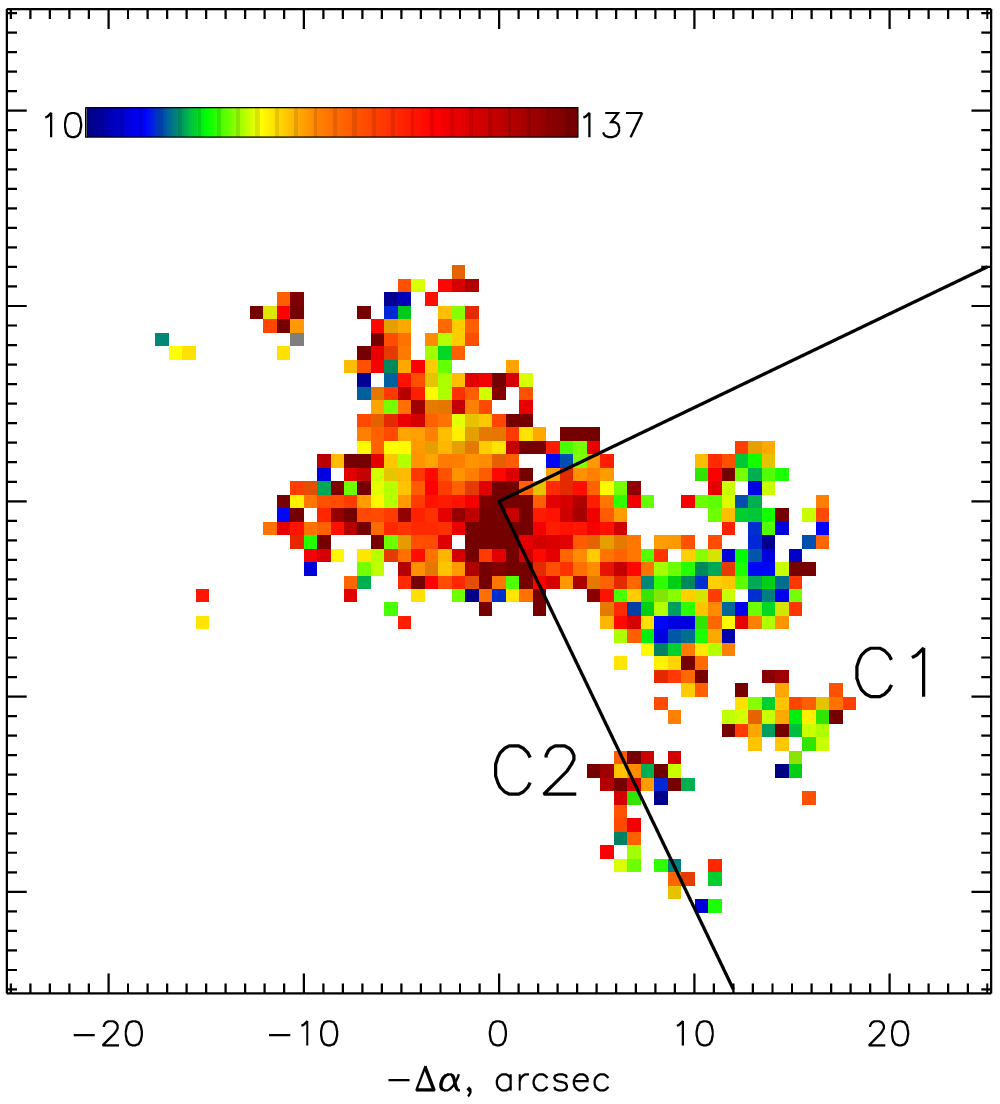}
}
\caption{Mrk~78 observations. Top left: the SCORPIO $R$-band image with the~long-slit position, a candidate  companion galaxy marked as `G'; top right: the map of the $[OIII]{\lambda}4959$ emission derived from the FPI data; bottom: the line-of-sight velocity field  in the [OIII] emission line (left) and the  velocity dispersion map with the marked ionization cone position (right). The colour bars are in the km\,s$^{-1}$, the external gaseous clouds are marked as `C1' and `C2'.}
\label{fields}
\end{figure}

The ionization state of the EELR inside  the galaxy disc ($r<10-15''$) was analysed in the papers cited above in Sec.~\ref{intr}. However, the FPI map also revealed  separate ionized-gas clouds (labeled as C1 and C2 in Fig.~\ref{fields}) at a distance of ${\sim}$15-20${\farcs}$ (${\sim}$12-16 kpc) southwest of the galaxy. The line-of-sight  velocity field shows that the cloud velocities lie in the velocities range of the disc gas (10900--11500~ km\,s$^{-1}$). Whereas the velocities of C1 and C2 by 50--200~ km\,s$^{-1}$ exceed the nearest S-W side of the galactic disc. This indicates that both clouds are moving in the Mrk~78 gravitational potential, but rotate out of the galaxy plane on retrograde orbits possibly.

To determine the C1 and C2 ionization state, the SCORPIO-2 long-slit spectra were obtained along PA=127$\degr$. The slit position has been selected according to the centers of both clouds and a neighbouring galaxy with an unknown spectroscopic redshift (see Sec.~\ref{back}). 

\begin{figure}[h]
\centerline{
\includegraphics[width=\textwidth]{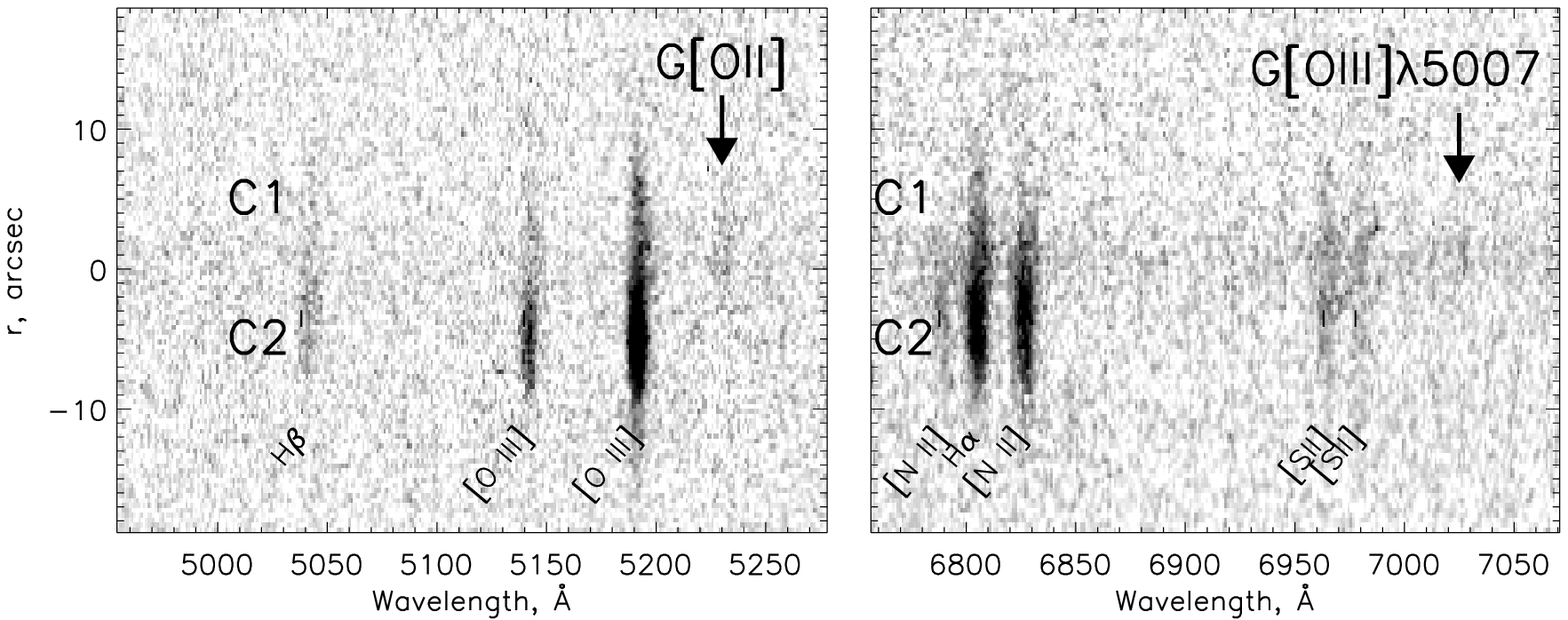}
}
\centerline{
\includegraphics[width=\textwidth]{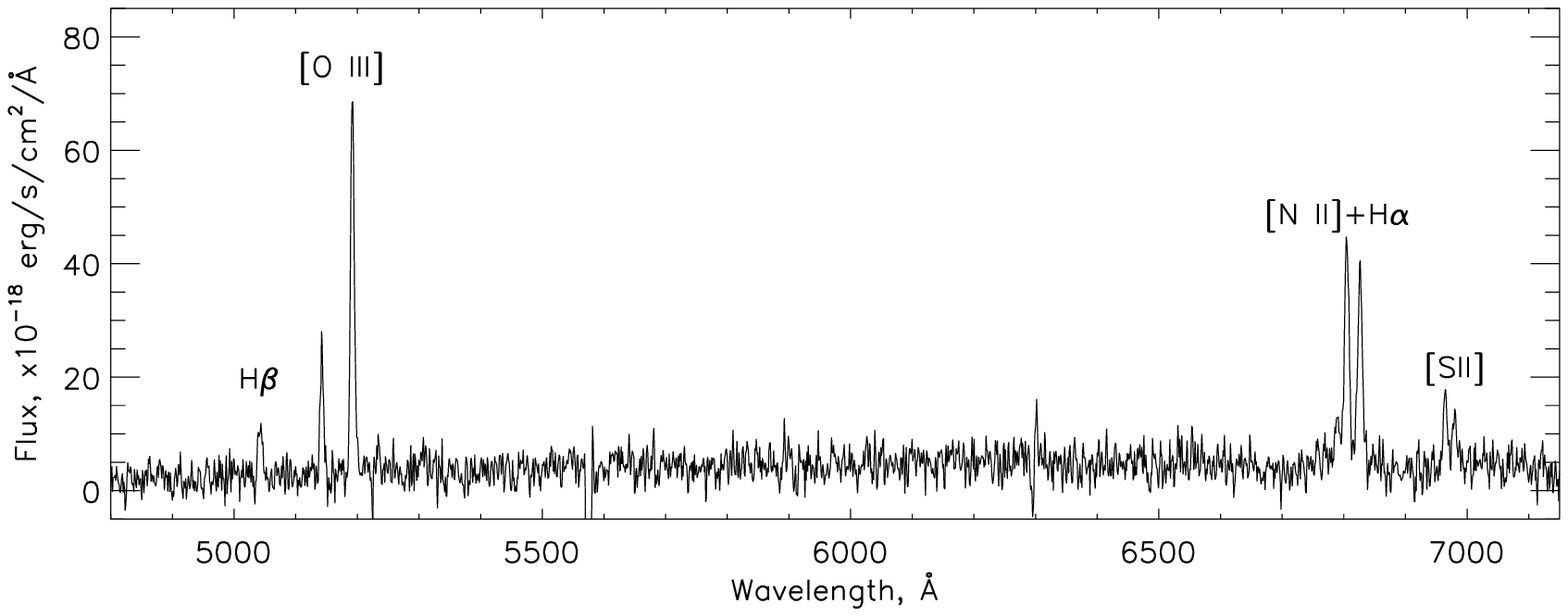}
}
\caption{Top: identification of the emission lines in the ionized-gas clouds (C1 and C2) and the nearby galaxy (G) long-slit spectra; bottom: the integrated spectrum of both clouds.}
\label{spec}
\end{figure}

To determine the ionization mechanism of the gas in C1 and C2, the emission-line ratio diagnostic diagrams \citep[BPT-diagrams after][]{1981PASP...93....5B} were used with the branches from \citet{2006MNRAS.372..961K} to separate different mechanisms of  gas excitation: the HII-regions and AGN (Seyfert or LINER). These diagrams involve   the following  line flux  ratios independent from the interstellar extinction: [N II]${\lambda}6583/H{\alpha}$,  [OIII]${\lambda}5007/H{\beta}$, and $[S II]{\lambda}6717+6731/H{\alpha}$. 

\begin{figure}[h]
\centerline{
\includegraphics[width=0.471\textwidth]{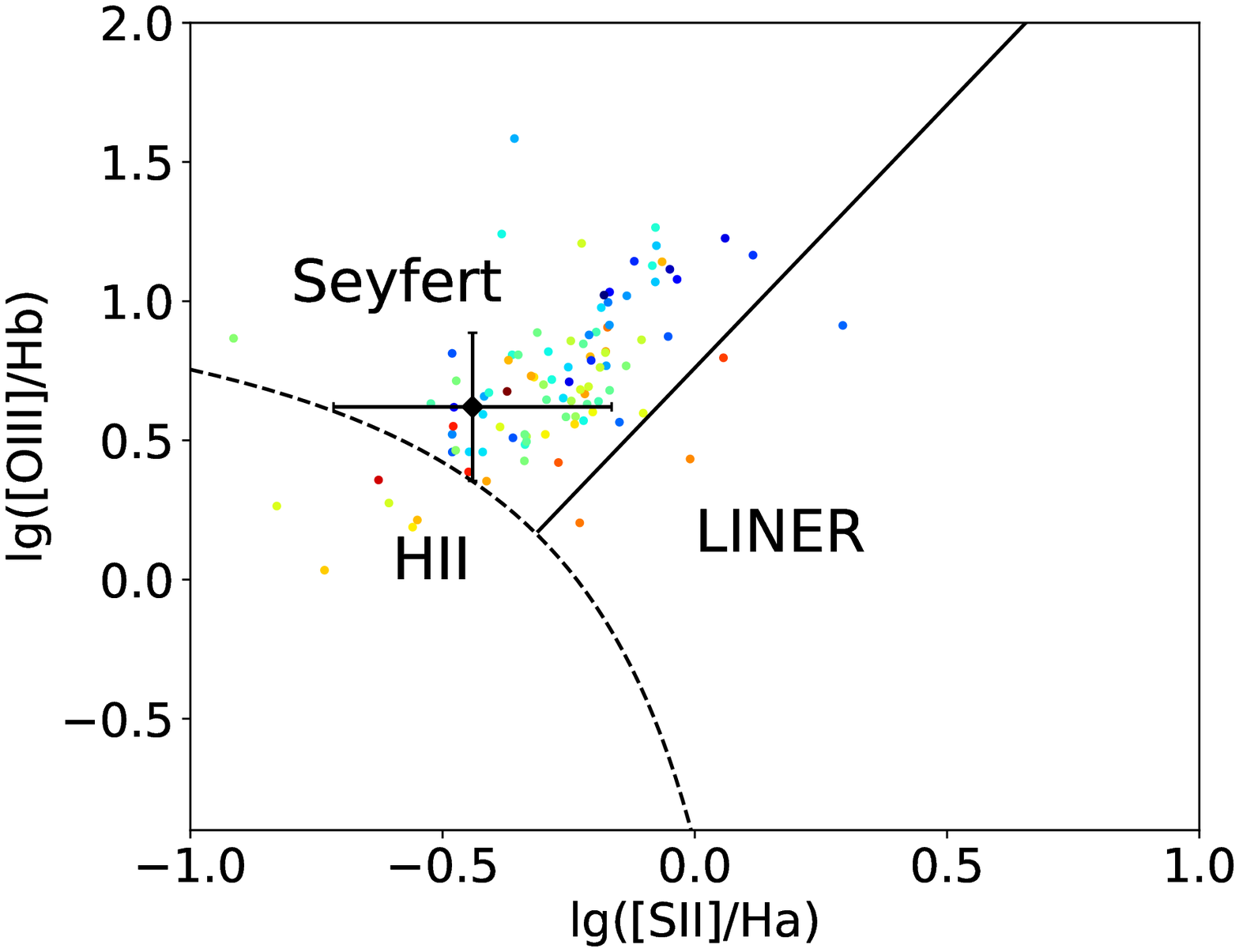}
\includegraphics[width=0.53\textwidth]{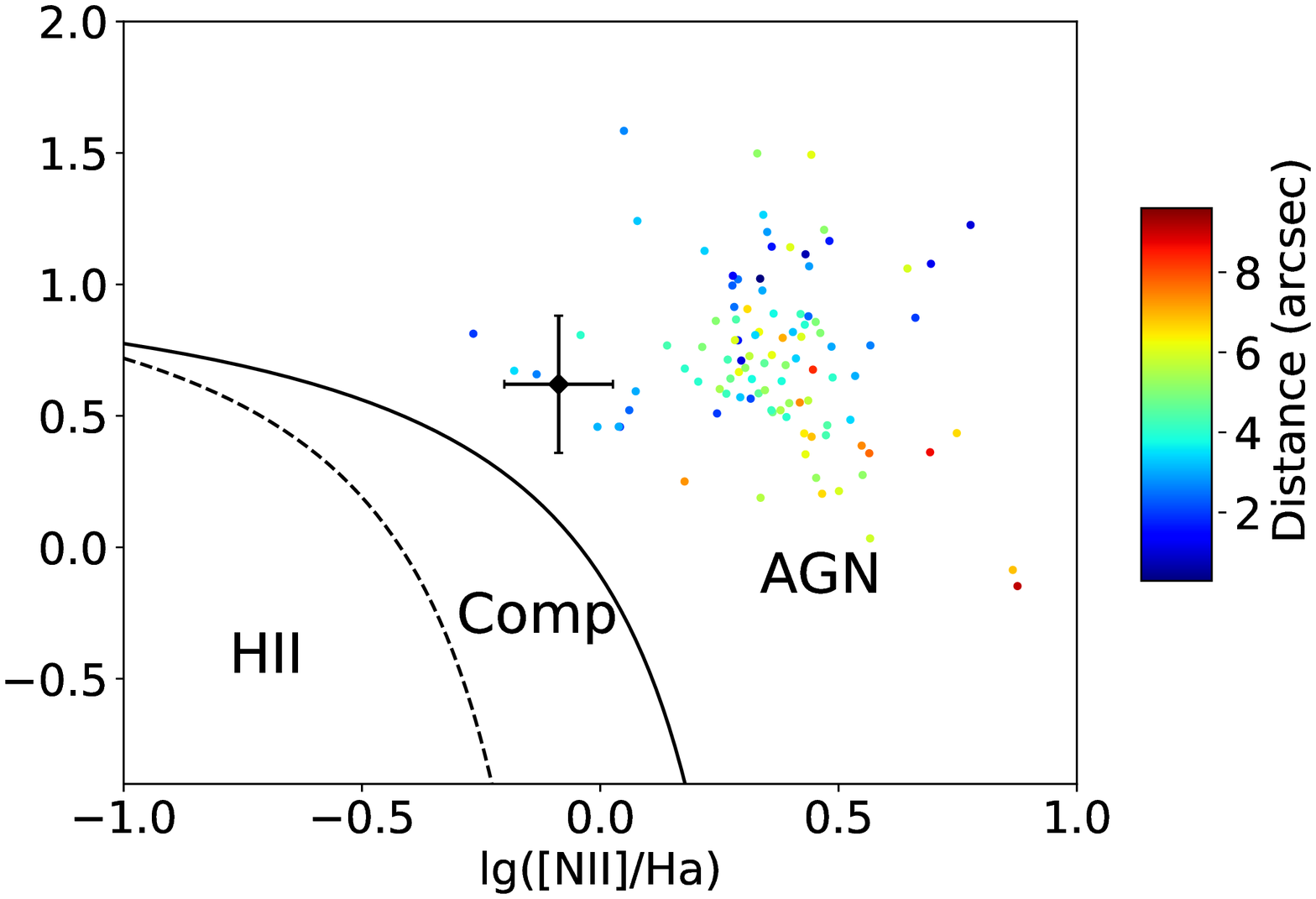}
}
\caption{BPT diagrams. The separate lines are taken from \citet{2006MNRAS.372..961K}. The circles denote the MPFS data for the circumnuclear region, the colour bar shows the projected distance of pixels on the MPFS maps from the nucleus. The diamond shows the line ratio of C1+C2 derived from the long-slit spectrum.}
\label{BPT}
\end{figure}

The integrated spectrum of the C1+C2 clouds is shown in Fig.~\ref{spec}, the BPT diagrams are presented in Fig. \ref{BPT}. We designated the EELR ionization mechanism according to the long-slit spectroscopy. To compare their ionization state to that of the circumnuclear region, the MPFS intergral-field spectroscopic data were used. It is clearly seen that the C1 and C2 ionization state is the same as that in the internal part of the EELR: the  Seyfert-type AGN radiation predominates. This result is in the agreement with the diagrams based on the  long-slit spectroscopy along the major galactic axis (PA=90$\degr$) presented in \cite{Keel2012}.

\cite{Fischer2011} estimated the parameters of the  ionized cones according to the HST  data for the inner ($r<4''$) brightest part of the emission-line structure. We overlaid the western side of this bicone (having an opening angle of ${\sim}$75${\degr}$) on the FPI dispersion map (Fig.~ \ref{fields}): C1 and C2 both fall into the projected cone borders. Therefore, the ionization state of the distant gas in the EELR is consistent with the model based on the analysis of the circumnuclear gas kinematics and ionization state. The ionized-gas  velocity dispersion in C1 and C2  (${40<\sigma<100}$~km\,s$^{-1}$, see Fig.~ \ref{fields})  is similar to the gas in the outer part of the galaxy disc and significantly smaller than that observed in the inner region, where jet-cloud interaction takes place. It means that the external part of the EELR is dynamically cold, i.e., the AGN highlight off-plane gas clouds are related  with the external accretion or tidal structures, but are not the result of the AGN outflow. The similar picture is observed in  most EELRs in Seyfert galaxies \citep[see] [and references therein]{Keel2015}.

\section{Possible dwarf companion}
\label{back}

Assumptions that Mrk~78 is a post-merging system were considered in a number of papers since the eighties of the last century. Morphological peculiarities on different spatial scales were considered as the galaxy interaction imprints: the nuclear dust lane \citep{2004AJ....127..606W},  the disturbed external isophotes in the S-W part of the galactic disc \citep{DeRobertis1987, Pedlar1989,  Keel2015}, the asymmetric distribution of the external [OIII] emission \citep{Unger1987,AfanasievSilchenko1991}.

\citet{DeRobertis1987} suggested that the disturbed external isophotes are caused by passing through a companion galaxy and the secondary emission of this region due to the companion galaxy nucleus, but the forthcoming HST data disproved this idea \citep{Fischer2011}. 

 The SDSS DR15 and SCORPIO $R$-band  images reveal the galaxy SDSS J074240.37+651021.4 with an unknown redshift  in the nearest neighbourhood of the C1 and C2 clouds at a distance of ${\sim}$18$''$ from the Mrk~78 nucleus (marked as `G' in  Fig.~\ref{fields}).  We have checked the possibility that this galaxy being a dwarf companion  of  Mrk~78   or remnant of a tidal structure. The obtained long-slit spectra (see the slit position in Fig.~\ref{fields}, PA=127$\degr$) show the continuum of  `G' with weak emission lines ($\lambda$=5230\AA~   and $\lambda$=7016\AA) marked with arrows in Fig.~\ref{spec}. We interpreted them as $[OII]{\lambda}3727$ and $[OIII]{\lambda}5007$ at the redshift $z=0.382$. In this case, the  $H{\beta}$ should be blended with  [S II]${\lambda}6731$ in the C1 and C2 clouds. This spectroscopic redshift is comparable to the photometric SDSS DR15 \citep{SDSSDR15} redshift: $z=0.367\pm 0.130$. Therefore, it is a distant background galaxy.

\section{Summary}
\label{sum}

The 6-m telescope observations   reveal a pair of the ionized-gas  clouds   southwest from the stellar disc of  Mrk~78 at the 12--16\,kpc projected distance from the AGN located beyond the galactic disc plane. The BPT diagrams  have demonstrated  that the AGN radiation makes the main contribution to the ionization state of these clouds. Moreover, the location of both clouds together with other external [OIII]-emitting structures in the west side of the Mrk~78 disc is in a good agreement  with the  ionization cone borders proposed by \cite{Fischer2011} according only to the circumnuclear region structure and kinematics.

The AGN outflow as a source of the described ionized-gas clouds   is the contrary to the dynamically cold gas in this part of the EELR. The most possible scenario is an external accretion, a minor merging, or a tidal interaction with the companion. The galaxy  SDSS~J074240.37+651021.4 at a projected distance of ${\sim}$\,18\,$''$ from the Mrk~78 nucleus in the southwest direction was   considered  a candidate. However, the  long-slit spectroscopy data gave a redshift of 0.382;  it is a distant background galaxy. Thereby, the origin of the gas in the outer  part of the Mrk~78 EELRs is still unclear.

\acknowledgements
This study was supported by the Russian Science Foundation, project no. 17-12-01335 `Ionized gas in galaxy discs and beyond the optical radius' and based on the observations conducted with the 6-m telescope of the Special Astrophysical Observatory of the Russian Academy of Sciences carried out with the financial support of the Ministry of Science and Higher Education of the Russian Federation.  Authors are  grateful 
to Victor Afanasiev for his interest to this work.
This research has made use of the NASA/IPAC Extragalactic Database (NED), which is operated by the Jet Propulsion Laboratory, California Institute of Technology, under contract with the National Aeronautics and Space Administration. 


\end{document}